\documentclass[groupedaddress,amsmath,amssymb,twocolumn, aps,prl]{revtex4}
\usepackage{graphicx}
\usepackage{amsmath}
\usepackage{amssymb}
\usepackage{braket}
\usepackage{dcolumn}
\usepackage[colorlinks=true,linkcolor=blue,citecolor=blue,urlcolor=blue]{hyperref}

\begin{document}

\title{General Construction of Quantum Error-Correcting Codes from Multiple Classical Codes }
\author{Yue Wu}
\affiliation{Institute for Advanced Study, Tsinghua University, Beijing, 100084, China}

\author{Meng-Yuan Li}
\affiliation{Institute for Advanced Study, Tsinghua University, Beijing, 100084, China}

\author{Chengshu Li}
\affiliation{Institute for Advanced Study, Tsinghua University, Beijing, 100084, China}

\author{Hui Zhai}
\email{hzhai@tsinghua.edu.cn}
\affiliation{Institute for Advanced Study, Tsinghua University, Beijing, 100084, China}
\date{\today}

\begin{abstract}
The hypergraph product (HGP) construction of quantum error-correcting codes (QECC) offers a general and explicit method for building a QECC from two classical codes, thereby paving the way for the discovery of good quantum low-density parity-check codes. In this letter, we propose a general and explicit construction recipe for QECCs from a total of $D$ classical codes for arbitrary $D$. Following this recipe guarantees the obtainment of a QECC within the stabilizer formalism and nearly exhausts all possible constructions. As examples, we demonstrate that our construction recovers the HGP construction when $D=2$ and leads to four distinct types of constructions for $D=3$, including a previously studied case as one of them. When the input classical codes are repetition codes, our $D=3$ constructions unify various three-dimensional lattice models into a single framework, encompassing the three-dimensional toric code model, a fracton model, and two other intriguing models not previously investigated. Among these, two types of constructions exhibit a trade-off between code distance and code dimension for a fixed number of qubits by adjusting the lengths of the different classical codes, and the optimal choice can simultaneously achieve relatively large values for both code distance and code dimension. Our general construction protocol provides another perspective for enriching the structure of QECCs and enables the exploration of richer possibilities for good codes.
\end{abstract}

\maketitle

Quantum superposition and entanglement endow quantum computers with exponentially enhanced computational power for specific problems compared to classical computers. However, these same properties render qubits extraordinarily vulnerable to environmental noise, making the realization of a practically useful quantum computer extremely challenging. The standard approach to overcoming this challenge is to protect quantum information using quantum error-correcting codes (QECCs) ~\cite{Nielsen2010,Knill1997,Terhal2015}. Since their initial constructions by Shor~\cite{Shor1995}, Steane~\cite{Steane1996a,Steane1996b}, and Kitaev~\cite{Kitaev1997} in the 1990s, QECCs have drawn inspiration from classical error-correcting codes, and these early quantum codes were later unified under the Calderbank--Shor--Steane (CSS)~\cite{Calderbank1996,Steane1996a,Steane1996b} and stabilizer formalisms~\cite{Gottesman1996,Gottesman1997}.

A significant breakthrough came in 2014 with the hypergraph product (HGP) code introduced by Tillich and Z\'emor~\cite{Tillich2014}. This construction provides a general and explicit recipe for building a quantum code from any two classical linear codes, where the properties of the resulting quantum code are determined by the parity-check matrices of the classical codes. The HGP code serves as a foundational method for constructing quantum low-density parity-check (qLDPC) codes, which can achieve constant rate and a code distance that grows as the square root of the number of physical qubits, allowing for arbitrary suppression of errors with increasing system size. Subsequent research building upon this line has led to the discovery of good qLDPC codes~\cite{Breuckmann2021, HastingsHaahODonnell2022,Panteleev2022,Leverrier2022}. Recent work also highlights that the HGP code can be straightforwardly implemented in atom array platforms by leveraging their reconfigurable hardware advantages~\cite{hengyunzhou2024,Pecorari2025}. 

However, most existing works have focused on constructing quantum codes from two classical codes. Only a few studies have considered the case of three or more classical codes as inputs, and these are neither as general nor as explicit as the HGP construction~\cite{BravyiHastings2014, Ostrev2024, Vedika, Zeng2019, Audoux2019, Leverrier20221}. This work aims to provide a general and explicit recipe for constructing QECC from multiple classical codes, thereby enabling systematic and nearly exhaustive exploration of all possibilities. This provides another direction to extend the structure of QECC, as well as topological models that they correspond to. 

\textit{Recipe of the General Construction.}  For the Tanner graph of each classical code $\mathcal{C}$, we denote its bits by $b = \{b_j| j = 1, \dots, L^\mathrm{b}\}$ and its checks by $c = \{c_i|i = 1, \dots, L^\mathrm{c}\}$. The check matrix $H$ is an $L^\mathrm{c} \times L^\mathrm{b}$ matrix that defines how the checks in $c$ act on the bits in $b$: a check $c_i$ acts on a bit $b_j$ if $H_{ij} \neq 0$. The dual code $\tilde{\mathcal{C}}$ effectively interchanges the roles of bits and checks, so that $b$ serves as checks and $c$ as bits, and correspondingly, the transpose of $H$ defines how $b$ acts on $c$.

For a total of $D$ classical codes $\mathcal{C}^l$ with $l = 1, \dots, D$, we introduce $2^D$ distinct blocks, each labeled by a $D$-tuple $(w^1, \dots, w^D)$ where each $w^l$ is either $b^l$ or $c^l$. Without loss of generality, we will assign some blocks with an even number of $b$'s to be qubits in the quantum code, and some blocks with an odd number of $b$'s to be either $X$- or $Z$-checks. The assignment follows the following recipe: 

\begin{enumerate}
  \item \textbf{Initialization with $Z$-check blocks:} Assign a set of blocks containing an odd number of $b$'s as $Z$-checks.
  \item \textbf{Generation of qubit blocks:} Define a \textit{FLIP} as an operation that interchanges $b$ and $c$ for a single classical code. Multiple \textit{FLIP}s can be applied simultaneously to different codes. Apply an odd number or several different odd numbers of \textit{FLIP}s to all $Z$-check blocks, and assign the resulting blocks as qubits.
  \item \textbf{Generation of $X$-check blocks:} Apply to all qubit blocks the same number of \textit{FLIP}s as in the previous step. Discard the resulting block if it already corresponds to a $Z$-check; Otherwise, assign it as an $X$-check.
\end{enumerate}

The check rule follows the same spirit as the HGP construction. In a given construction, we first specify the rule that determines how a check block acts on a qubit block based on the number of \textit{FLIP}s by which they differ. For instance, if the rule permits only a single \textit{FLIP} difference, and the check block and qubit block differ in the $l$-th component of the $D$-tuple, specifically, if the $l$-th entry is $c^l$ in the check block and $b^l$ in the qubit block, then the check acts on the qubit via the matrix $H^l$. Conversely, if the $l$-th entry is $b^l$ in the check block and $c^l$ in the qubit block, the check is determined by $(H^l)^\text{T}$. Next, suppose the rule allows them to differ by multiple \textit{FLIP}s. In that case, the overall check rule is given by the joint product of the corresponding matrices (either $H^l$ or its transpose) over all differing components.

It can be verified that all QECCs constructed via this procedure are stabilizer codes. The detailed proof is provided in the End Matter. In what follows, we present several illustrative examples.

\begin{figure}[t]
    \centering
    \includegraphics[width=\columnwidth]{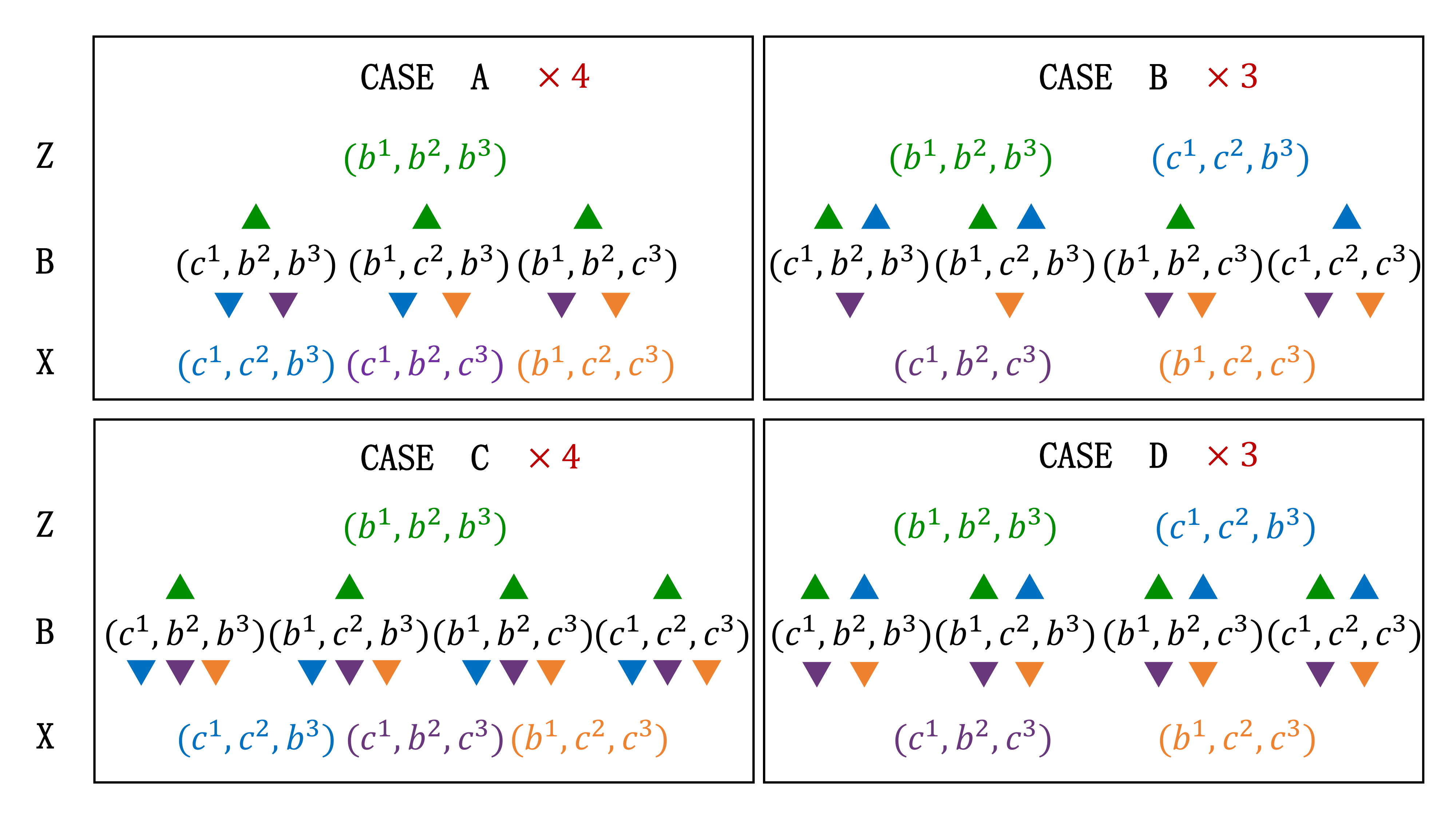}
    \caption{All four types of quantum codes constructed from three classical codes are presented. A representative example is shown for each type, with the total number of distinct codes of that type indicated by ``$\times n$''. The first and third rows correspond to $Z$-check and $X$-check blocks, respectively, while the middle row contains all qubit blocks. The upward- and downward-pointing triangles of different colors adjacent to each block indicate that the block is checked by the $Z$- and $X$-check blocks of the corresponding color.  }
    \label{3-d-code}
\end{figure} 

\textit{Revisit of $D=2$.} For $D=2$, our construction reduces to the HGP construction, which is the only construction for combining two classical codes. Nevertheless, it is instructive to revisit this basic case to familiarize readers with our protocol. In Step 1, we start with $(b^1, c^2)$ as the $Z$-check block. For $D=2$, only a single \textit{FLIP} is allowed, which either converts $b^1$ to $c^1$ or $c^2$ to $b^2$, yielding $(c^1, c^2)$ and $(b^1, b^2)$ as qubit blocks. Finally, applying a single \textit{FLIP} again to each qubit block gives $(c^1, b^2)$ as the $X$-check block.

\textit{Classification of $D=3$.} Having reviewed the case of $D=2$, we now turn to the construction with three classical codes. The corresponding results are summarized in Fig.~\ref{3-d-code}.

For case A, we begin with a single block, say $(b^1, b^2, b^3)$, as the $Z$-check. There are four distinct choices of blocks with an odd number of $b$'s to start with, corresponding to four different quantum codes of this type. We define the check rule such that a check block and a qubit block differ by exactly one \textit{FLIP}. Applying a single \textit{FLIP} to the $Z$-check block in each of the three possible positions generates the qubit blocks $(c^1, b^2, b^3)$, $(b^1, c^2, b^3)$, and $(b^1, b^2, c^3)$. The $Z$-check acts on all three of these qubit blocks. Subsequently, applying \textit{FLIP} again to each qubit block yields the $X$-check blocks, which act on a qubit block if they differ from it by exactly one \textit{FLIP}. We note that this construction coincides with that presented in Ref. \cite{Zeng2019} using the chain complex method.

For case B, we begin by selecting two blocks as $Z$-checks, with six distinct choices available. However, codes that differ only by an exchange of $X$- and $Z$-checks are considered equivalent, which reduces the number of distinct choices from six to three. The remaining steps follow the same procedure as in case A. We note that this construction can be interpreted as first taking the classical product of two codes, followed by the HGP construction with the third code. In the example shown in Fig.~\ref{3-d-code}, this corresponds to the HGP of $\mathcal{C}^1 \times \mathcal{C}^2$ and $\mathcal{C}^3$.

Cases C and D share the same initial step as cases A and B, respectively, but differ in the check rule. Here, we allow a check block to act on a qubit block if they differ by either one \textit{FLIP} or three \textit{FLIP}s. For example, the $Z$-check block $(b^1, b^2, b^3)$ can check the qubit block $(c^1, c^2, c^3)$, and the $X$-check block $(b^1, c^2, c^3)$ can check the qubit block $(c^1, b^2, b^3)$. To the best of our knowledge, these two QECCs have not been previously discussed. 

If we instead initiate Step 1 with three blocks, the resulting codes are equivalent to case A or C under the exchange of $X$- and $Z$-checks. Therefore, all these four types of codes exhaust all non-trivial constructions from three classical codes, provided we exclude cases where the resulting quantum code can be decoupled into lower-dimensional codes, or where a check block acts on different qubit blocks as different types of checks~\cite{Vijay2016, Vedika}. This construction recipe can be directly extended to larger values of $D$ following the same principles, although the resulting code structures become increasingly complex.

\begin{figure}[t]
    \centering
    \includegraphics[width=\columnwidth]{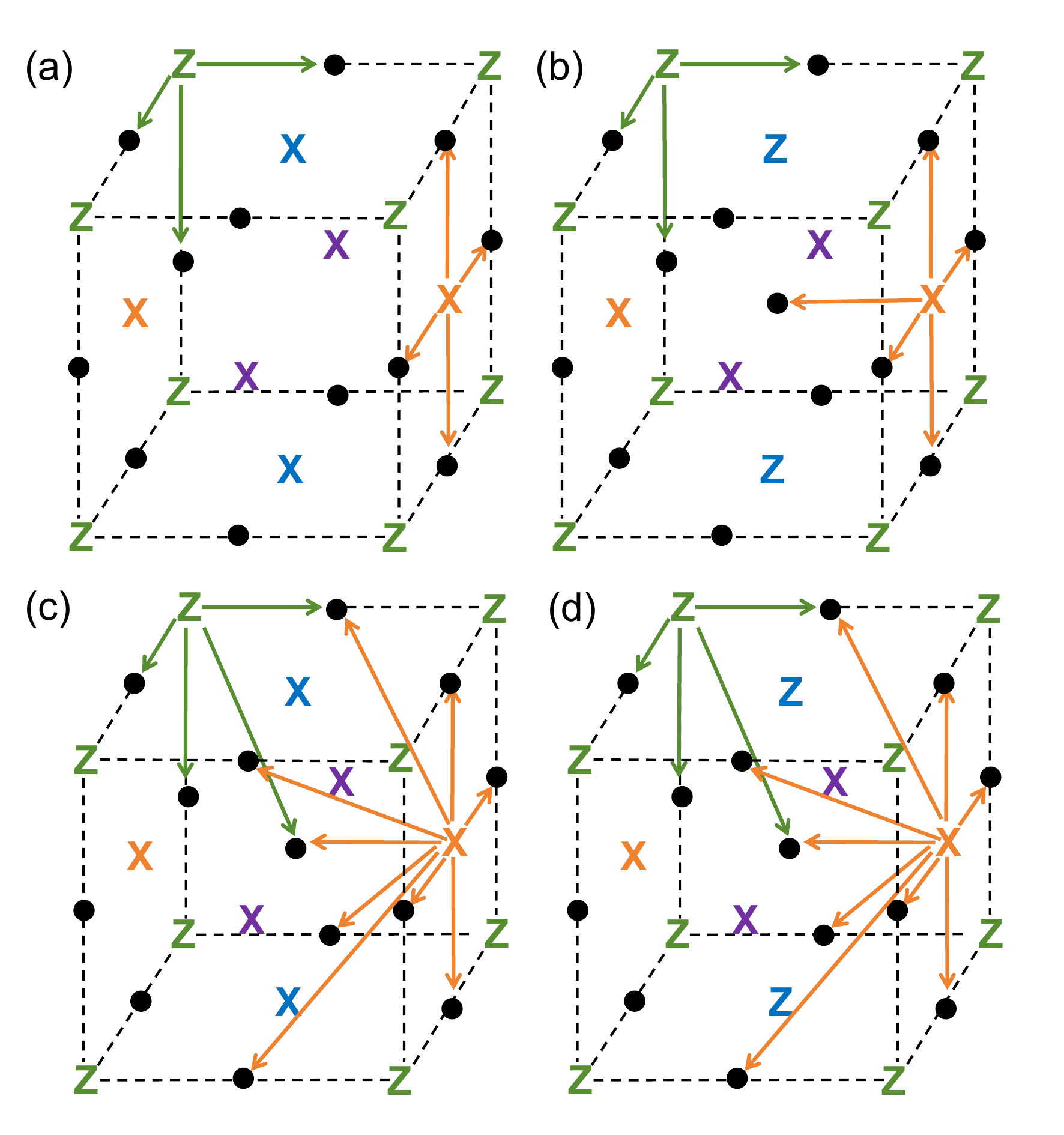}
    \caption{The unit cells of the four distinct lattice models corresponding to the four constructions in the $D=3$ case shown in Fig.~\ref{3-d-code}. Black dots represent qubits. $Z$ and $X$ symbols in different colors denote $Z$-checks and $X$-checks, with colors matching those in Fig.~\ref{3-d-code}. Green and yellow arrows indicate the action of each check by connecting it to the qubits on which it acts.
     }
    \label{models}
\end{figure}

\textit{Input as Repetition Code.} Below, to develop intuition about the properties of the different constructions, we consider the simplest scenario where all input classical codes are repetition codes. Note that in a classical repetition code, each check connects two neighboring bits. Therefore, in the case of $D=3$, if a check block and a qubit block are connected by a single \textit{FLIP}, the check acts on two qubits in that block; if they are connected by three \textit{FLIP}s, the check applies to $2^3 = 8$ qubits in the block. For the four different constructions shown in Fig.~\ref{3-d-code}, the corresponding three-dimensional models for each are presented in Fig.~\ref{models}.

In these models, the check from the block $(b^1,b^2,b^3)$ (green) is placed at the corner of each cube, while the checks from the other three blocks (blue, purple, and yellow) are positioned at the center of each plaquette. The qubits from the blocks $(b^1,b^2,c^3)$, $(b^1,c^2,b^3)$, and $(c^1,b^2,b^3)$ are located at the midpoints of the edges, and the qubits from the block $(c^1,c^2,c^3)$ are placed at the body center of each cube.

In case A, only one type of $Z$-check exists, located at the cube corners, each acting on the six qubits situated along the connected edges. The $X$-checks reside at the center of each plaquette, with each check acting on the four qubits located along its edges. This structure corresponds precisely to the three-dimensional generalization of the toric code model~\cite{3dtoric}.

Moving from case A to case B, a pair of $X$-checks on the horizontal plaquettes is replaced by a pair of $Z$-checks, and additional qubits are introduced at the cube centers. Each $Z$-check continues to act on six qubits, as in case A, but each $X$-check now also involves the extra qubits located at the centers of the two cubes adjacent to the plaquette. As a result, each $X$-check also acts on a total of six qubits in this configuration. This model represents a $\pi/4$-rotated version of a fracton model studied previously~\cite{Fuji, Liu-Ji, Chenxie}.

Moving from case A to case C, the structure of checks remains unchanged, but additional qubits are introduced at the cube centers, as in case B. Unlike in case B, each $Z$-check now also acts on the cube-center qubits. Specifically, each $Z$-check involves the center qubit of each of the eight cubes sharing that corner, bringing the total number of qubits each $Z$-check acts on to fourteen. Similarly, each $X$-check not only acts on the cube-center qubits but also on the qubits located along the eight edges perpendicular to and attached to the plaquette, also increasing the total number of qubits it acts on to fourteen.

Moving from case C to case D is more straightforward: it involves replacing two $X$-checks on the horizontal plaquettes with $Z$-checks, while keeping their action rules unchanged. To the best of our knowledge, neither of the models derived from cases C and D has been previously studied as a lattice model.

\begin{table}[t]
\begin{tabular}{|l|l|l|}
\hline
 &  \textbf{k} & \textbf{d} \\
\hline
A   & $3$ & $\min(L_1,L_2,L_3)$ \\
\hline
B  & $4\gcd(L_1,L_2)$ & $\min(2\mathrm{lcm}(L_1,L_2),L_1L_2,L_3)$ \\
\hline
C   & $\text{--}$ & $\min(L_1,L_2,L_3, \beta)$ \\
\hline
D  & $4\gcd(L_1,L_2)+\alpha(L_3-1)$ & $\min(2\mathrm{lcm}(L_1,L_2),L_1L_2,L_3)$ \\
\hline
\end{tabular}
\caption{Summary of general expressions for the code dimension $k$ and code distance $d$ for the four constructions in $D=3$, with the classical codes being repetition codes. Here, $\mathrm{gcd}$ denotes the greatest common divisor, and $\mathrm{lcm}$ denotes the least common multiple. For case C, $\beta=5$ if $L_{1,2,3}$ are pairwise mutually prime; otherwise $\beta=4$. For case D, $\alpha=8$ if 3 divides both $L_1$ and $L_2$; otherwise $\alpha=0$.
}
\label{k-d-table}
\end{table}

\begin{figure}[t]
    \centering
    \includegraphics[width=\columnwidth]{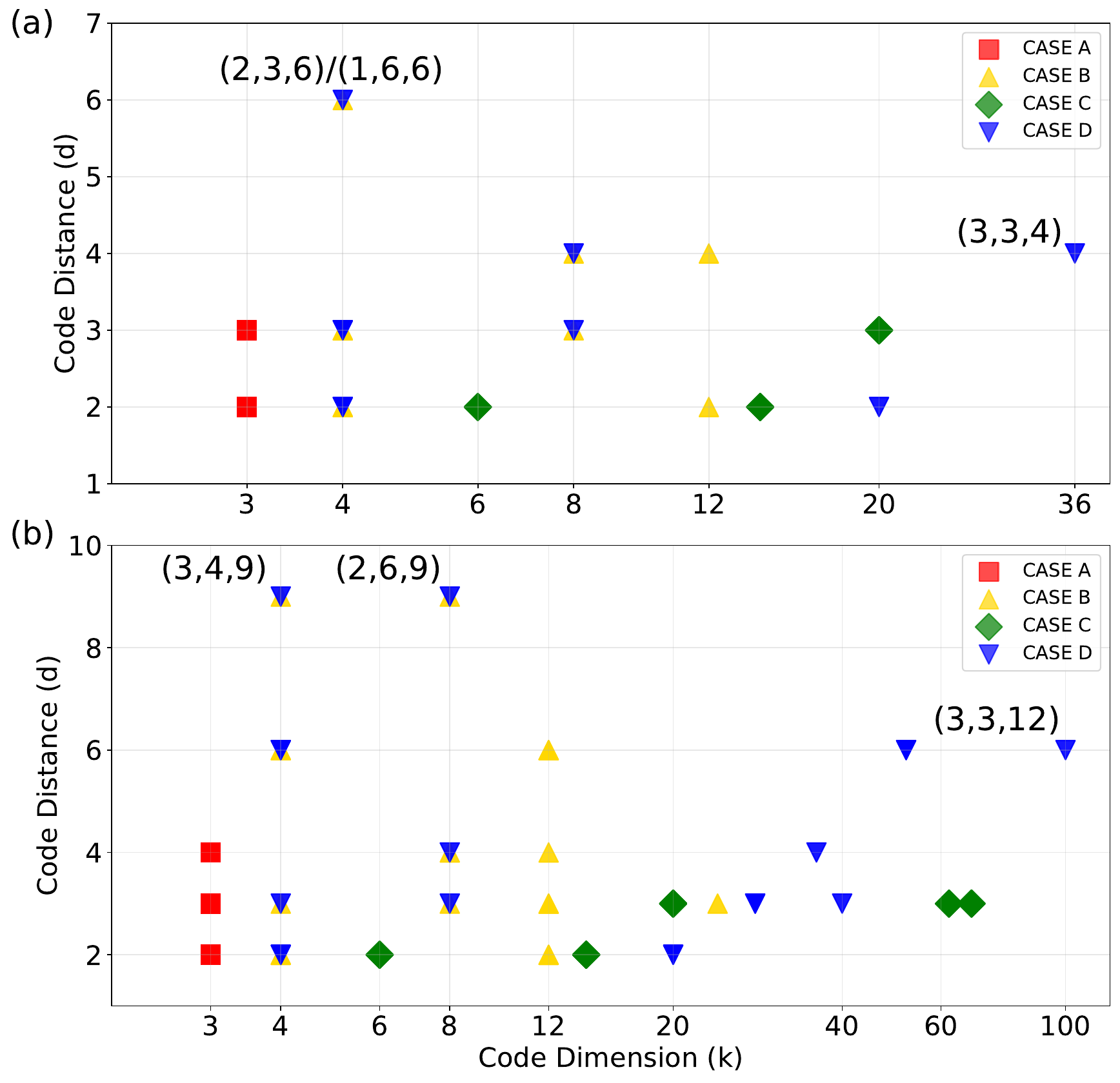}
    \caption{ The code dimension $k$ and code distance $d$ for four different kinds of codes at $D=3$. Here, the total number of qubits is fixed at $n=144$ for (a) and $n=432$ for (b). For a fixed $n$, various combinations of $(L_1,L_2,L_3)$ are possible, with each combination corresponding to a distinct data point for a given case. The number $(L_1,L_2,L_3)$ aside the data point represents the combination with which the largest $k$ or the largest $d$ is achieved. 
     }
    \label{k-d-figure}
\end{figure}

We consider three repetition codes of lengths $L^b_i=L^c_i=L_i$ ($i=1,2,3$). The total number of qubits is $n=3L_1L_2L_3$ for case A and $n=4L_1L_2L_3$ for the other three cases. The number of logical qubits, i.e., the code dimension $k$, corresponds to the ground-state degeneracy of the lattice model in Fig.~\ref{models} under periodic boundary conditions along all three directions. The number of independent checks can be determined either by analytical enumeration~\cite{supple} or by numerically evaluating the rank and kernel of the check matrices. The value of $k$ is then obtained by subtracting the number of independent checks from the total number of qubits $n$. Analytical enumeration yields the expressions summarized in Table~\ref{k-d-table} except for case C, as confirmed numerically in Fig.~\ref{k-d-figure}.

For a fixed total number of qubits, different choices of $(L_1, L_2, L_3)$ generally lead to different values of $k$ and $d$. However, in case A, $k = 3$ remains constant across all such choices, reflecting a fixed ground-state degeneracy consistent with the topological order of the three-dimensional toric code model. In case B, $k$ depends on the greatest common divisor of $L_1$ and $L_2$, and is independent of $L_3$. As seen in Fig.~\ref{models}(b), the vertical direction differs from the horizontal ones: the plaquette centers are occupied by $Z$-checks instead of $X$-checks. This structure aligns with a type of fracton order, in which excitations are constrained to move along the third spatial direction~\cite{Chamon2005,Haah2011,VijayHaahFu2015,Nandkishore2019}. For case C, we have not yet arrived at an analytical expression for $k$; however, numerical results in Fig.~\ref{k-d-figure} indicate that $k$ is non-constant and can reach relatively large values. Case D shares the same $k$ as case B when $L_1$ and $L_2$ are not both multiples of three, suggesting that it may also realize a fracton order. However, when both $L_1$ and $L_2$ are multiples of three, $k$ acquires an extra term that grows linearly with $L_3$, implying a substantially larger ground-state degeneracy for the lattice model. This feature is also reminiscent of recent discussions of the bivariate bicycle code~\cite{BB,chen2025}, making the model particularly interesting for future study.

The code distance is determined by identifying the minimum length of logical operators, either through analytical derivation~\cite{supple} or numerical probabilistic algorithm~\cite{numerical_method}. In case A, $d$ follows the standard behavior of the toric code, increasing linearly with the length of the classical repetition code. Cases B and D share the same expression for $d$, exhibiting strong anisotropy across different directions. In contrast, case C does not yield a good code in terms of distance, as $d$ is always bounded by $d \leqslant 5$. This limitation arises from the presence of local logical operators in the code.

Thus, case A maintains a constant $k=3$, while case C is always limited to $d \leqslant 5$, regardless of the total number of qubits. In contrast, cases B and D exhibit a trade-off between code dimension $k$ and code distance $d$. Specifically, for a given $n$, different combinations of $(L_1, L_2, L_3)$ allow one to either target the maximum possible $d$ or aim for the largest $k$ with a relatively smaller $d$. For example, with $n=144$ as shown in Fig.~\ref{k-d-figure}(a), the maximum $d=6$ (with $k=4$) is achieved using $(L_1,L_2,L_3)=(1,6,6)$ or $(2,3,6)$ under case B or D, while the largest $k=36$ (with $d=4$) corresponds to $(3,3,4)$ under the case D. Similarly, for $n=432$ in Fig.~\ref{k-d-figure}(b), the combination $(2,6,9)$ yields the maximum $d=9$ (with $k=8$) under case B or D, and $(3,3,12)$ gives the largest $k=100$ (with $d=6$) under case D.

\textit{Outlook.} In summary, our general construction principle provides a systematic approach to building high-dimensional QECCs within the stabilizer formalism. We demonstrate that, in three dimensions and with repetition codes as inputs, this framework already unifies four distinct types of lattice models, two of which have not been previously studied. It would be intriguing to employ this construction to uncover more interesting models in even higher dimensions. Our results also reveal that varying the lengths of different classical codes can achieve significant tunability in both the code distance and the code dimension, even with a fixed number of qubits. An optimal choice leads to relatively large values for both parameters. This construction principle thus enables the exploration of a far richer structure of QECCs. This construction also maintains the block structure, which is suitable for implementation on the reconfigurable atom array platform as HGP codes. One of the most important future directions is to investigate whether certain constructions can exhibit even better properties when used with low-density parity-check codes as classical inputs.

\textit{Acknowledgment.} We thank Liang Mao, Shang Liu, Yifei Wang, Yu-An Chen, and Hao Song for helpful discussions. This work is supported by the Shanghai Committee of Science and Technology grant No. 25LZ2600800 and National Natural Science Foundation of China No. 12488301(H.Z.), No. U23A6004 (H.Z.), and No. 12504307 (C.L.). C.L. is also supported by the Tsinghua University Dushi program. 

\textit{Code availability.} The codes for this work are available at
https://github.com/May0703/General-Construction-of-Quantum-Error-Correcting-Codes-from-Multiple-Classical.

\textit{End Matter.}
In this section, we present a rigorous proof that the general construction introduced in the main text produces a valid QECC within the stabilizer formalism. The key requirement that all $X$-stabilizers commute with all $Z$-stabilizers is verified. The commutativity condition is equivalent to requiring that the supports of any $X$-check and any $Z$-check overlap on an even number of qubits. 

Within our framework, a $D$-tuple $(w^1, \dots, w^D)$, where each entry $w^l$ is either $b^l$ or $c^l$, represents a group of stabilizers or qubits. Each specific stabilizer or qubit is uniquely identified by an indexed $D$-tuple $(w^1_{\eta_1}, \dots, w^D_{\eta_D})$, with each $w^l_{\eta_l}$ chosen as either $b^l_{i_l}$ or $c^l_{j_l}$ from the $l$-th classical code, thereby specifying the exact bit or check selected at each position. 

Consider a general pair of $\hat{X}$-check and $\hat{Z}$-check, and denote the set of qubits on which they both act as $\Sigma$. We first introduce a mapping $f$ on $\Sigma$. It can be shown that this function possesses the following three properties: i) For any qubit $Q \in \Sigma$, $f(Q) \in \Sigma$; ii) $f(Q) \neq Q$ for any $Q \in \Sigma$; iii) $f$ is single-valued and reversible, and thus a one-to-one mapping.
Based on properties (i-iii), all elements in $\Sigma$ can be grouped into pairs, implying the total number of elements in $\Sigma$ must be even. This naturally leads to the conclusion that any $X$-check and $Z$-check overlap on an even number of qubits.

\begin{figure}[t]
    \centering
    \includegraphics[width=\columnwidth]{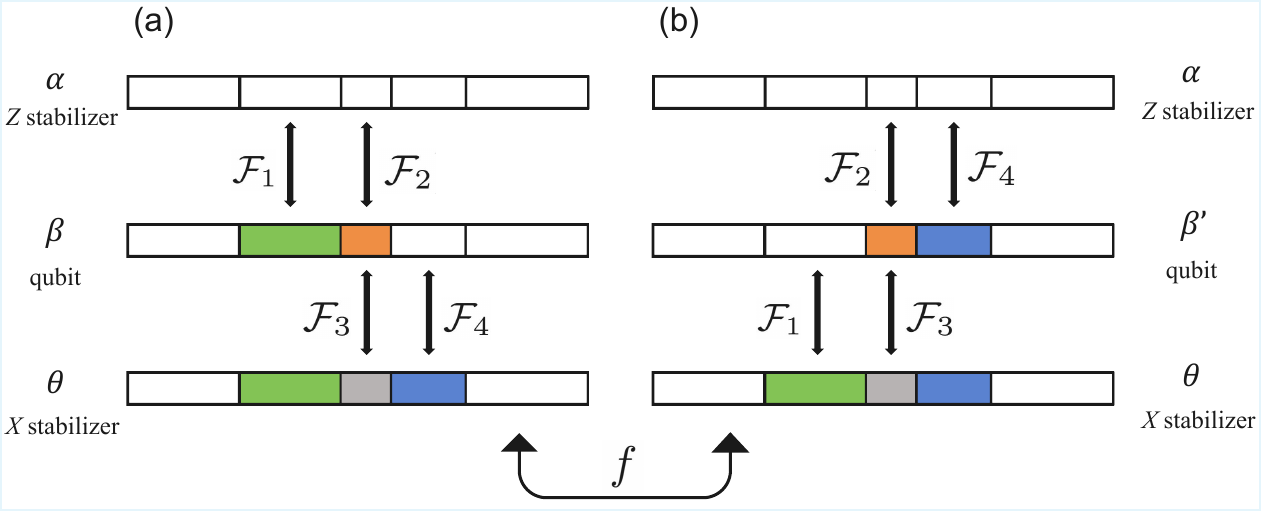}
    \caption{ Illustration of the \textit{FLIP}s and the induced pairing structure between qubits.
(a) The indexed $D$-tuples $\alpha$, $\beta$, and $\theta$ are depicted as colored bars where same color indicates same choices of indices at that position. They represent $Z$-stabilizer, qubit and $X$-stabilizer, respectively. $\beta$ is generated from $\alpha$ by applying $\mathcal{F}_1$ and $\mathcal{F}_2$, yielding the green and orange regions where $\beta$ differs from $\alpha$. It can also be generated by applying $\mathcal{F}_3$ and $\mathcal{F}_4$ to the $X$-stabilizer $\theta$.
(b) $\beta'$ is constructed from $\alpha$ using $\mathcal{F}_2$ and $\mathcal{F}_4$, or constructed from $\theta$ using $\mathcal{F}_1$ and $\mathcal{F}_3$.
The arrow below indicates that panels (a) and (b) are related by interchanging the order of 
$\mathcal{F}_1$ and $\mathcal{F}_4$. 
For visual clarity, the colored segments are drawn contiguously, although contiguity is not required in general. }
    \label{endmatter}
\end{figure}

We first define $f$. Let the $Z$-check, qubit-$Q$ and $X$-check be labeled by indexed $D$-tuples $\alpha$, $\beta$ and $\theta$, and we examine how the three tuples change to introduce four \textit{FLIPs}, see Fig.~\ref{endmatter}(a). If a sector of the tuple changes between $\alpha$ and $\beta$ but remain fixed between $\beta$ and $\theta$, the \textit{FLIP} that relates $\alpha$ and $\beta$ on this sector is labeled $\mathcal{F}_1$. Conversely, if a sector of the tuple changes between $\beta$ and $\theta$ but remains fixed between $\alpha$ and $\beta$, the corresponding \textit{FLIP} is labeled $\mathcal{F}_4$. Finally, if a sector of the tuple changes both between $\alpha$ and $\beta$ and between $\beta$ and $\theta$, we call the \textit{FLIP} between $\alpha$ and $\beta$ $\mathcal{F}_2$ and the \textit{FLIP} between $\beta$ and $\theta$ $\mathcal{F}_3$, both of which have support on this common sector. With the four \textit{FLIPs} in hand, we can define $f(Q)$ as the qubit from simply interchanging $\mathcal{F}_1$ and $\mathcal{F}_4$, which is labeled by the indexed $D$-tuple $\beta'$ in Fig.~\ref{endmatter}(b). Then $f(Q)$ is the the qubit obtained by applying $\mathcal{F}_4$ and $\mathcal{F}_2$ instead of $\mathcal{F}_1$ and $\mathcal{F}_2$ to $\alpha$. 

Now we prove the three properties of $f$. For property (i),
from the definition above we see that $f(Q)$ is checked by the $Z$ stabilizer $\alpha$, and what we need to verify is that $f(Q)$ is also checked by the $X$ stabilizer $\theta$. To see this, we note that $\mathcal{F}_1$ and $\mathcal{F}_4$ have non-overlapping support so they commute. For the same reason, $\mathcal{F}_1$ and $\mathcal{F}_4$ commute with $\mathcal{F}_2$ and $\mathcal{F}_3$. Therefore, $(\mathcal{F}_1\mathcal{F}_3)(\mathcal{F}_2\mathcal{F}_4)\alpha=(\mathcal{F}_4\mathcal{F}_3)(\mathcal{F}_2\mathcal{F}_1)\alpha=\theta$. So $f(Q)$ is indeed checked by the $X$ stabilizer $\theta$.

For (ii),
we note that $\mathcal{F}_1$ and $\mathcal{F}_4$ cannot be both empty. 
In that case, the indexed $D$-tuple $\alpha=(w^1_{\eta_1}, \dots, w^D_{\eta_D})$ is mapped to $\gamma=(w^1_{\eta_1'}, \dots, w^D_{\eta_D'})$, with each entry both $b$ or $c$. Hence, $\alpha$ and $\gamma$ belong to the same block, which is not allowed in our construction (cf. Fig.~\ref{3-d-code}). Therefore, we see that $f(Q)\neq Q$. We also note that $\mathcal{F}_2$ and $\mathcal{F}_3$ could be empty, which does not affect our deduction. 

For (iii), since $\mathcal{F}_1$ and $\mathcal{F}_4$ are fixed by the choice of $\alpha$, $\beta$, and $\theta$, our explicit construction involves no arbitrary choice and is therefore single-valued. Applying $f$ to $f(Q)$ again interchanges $\mathcal{F}_1$ and $\mathcal{F}_4$ once more, which returns $Q$; hence $f$ is reversible. The verification of conditions (i-iii) then concludes the proof.

\end{document}


\title{Supplementary Material for General Construction of Quantum Error-Correcting Codes from Multiple Classical Codes}
\author{Yue Wu}
\affiliation{Institute for Advanced Study, Tsinghua University, Beijing, 100084, China}

\author{Meng-Yuan Li}
\affiliation{Institute for Advanced Study, Tsinghua University, Beijing, 100084, China}

\author{Chengshu Li}
\affiliation{Institute for Advanced Study, Tsinghua University, Beijing, 100084, China}

\author{Hui Zhai}
\email{hzhai@tsinghua.edu.cn}
\affiliation{Institute for Advanced Study, Tsinghua University, Beijing, 100084, China}
\date{\today}

\maketitle

This supplementary material presents the analytical expressions for the code dimension and code distance corresponding to the four cases arising from different three-dimensional (3D) products of repetition codes, as discussed in the main text. The code dimension is determined by calculating the number of independent checks, which equals the total number of checks minus the number of dependent relations (i.e., redundancies) among them. The code distance is obtained by identifying all logical operators that commute with the checks and are not products of the stabilizers; the minimum weight among such operators defines the code distance.

\subsection*{Case A}
This case corresponds to the well-known 3D toric code. The $Z$-checks are defined on the corners of the cubes, with each check acting on the six qubits located on the connected edges.  Multiplying all $Z$-checks across the lattice results in the identity. Crucially, this constitutes the only such relation among the $Z$-checks, implying that the number of independent $Z$-checks is given by $L_1 L_2 L_3 - 1$, where $L_1$, $L_2$, and $L_3$ represent the lengths of the three constituent repetition codes.

\begin{figure}[t]
    \centering
    \includegraphics[width=\columnwidth]{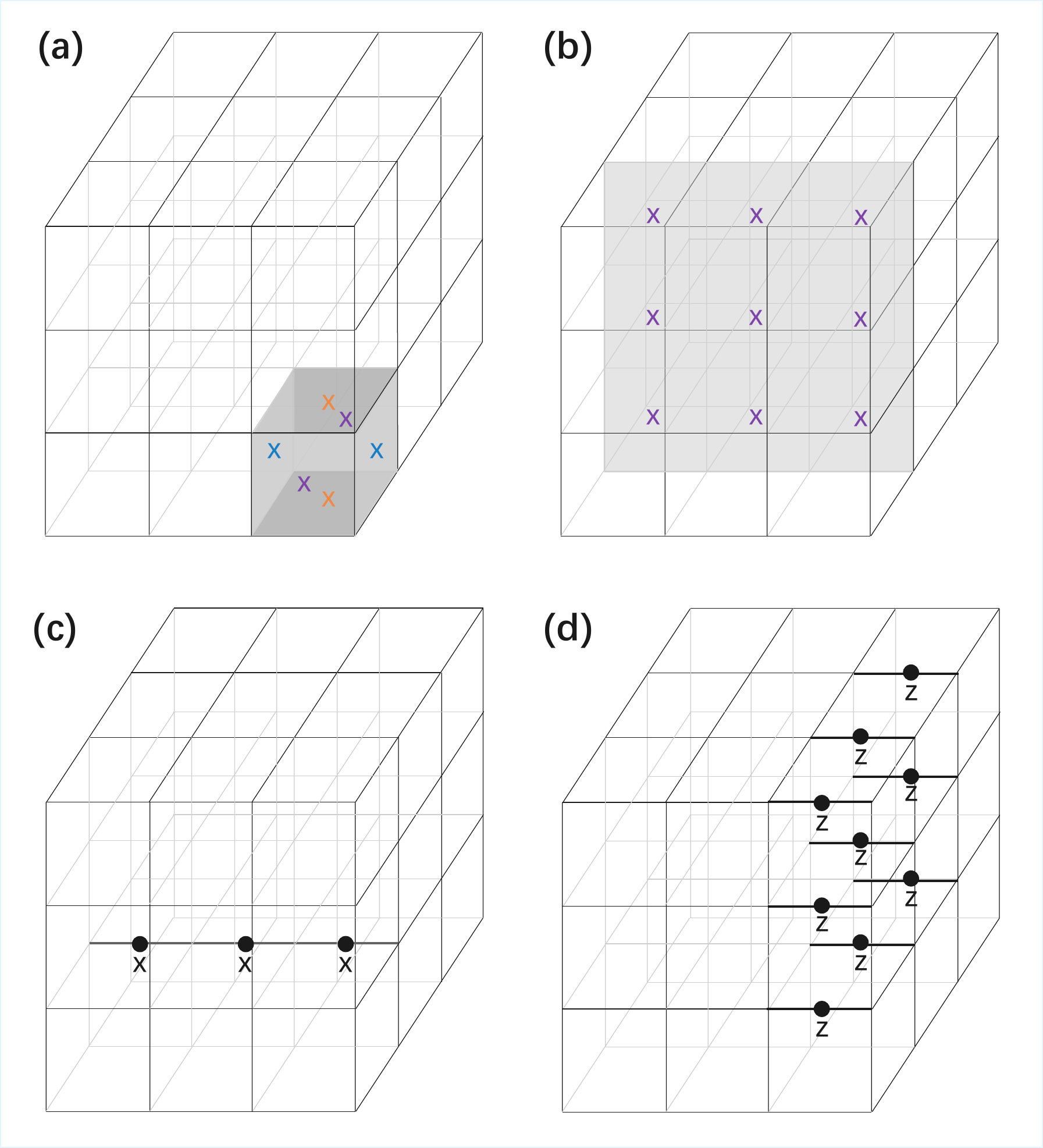}
    \caption{Redundancies of checks and logical operators in Case A. (a) and (b) Typical redundancies among the $X$-check operators, where the shaded faces indicate the participation of the $X$-checks located on them. (c) A representative logical $X$ operator. (d) A representative logical $Z$ operator.}
    \label{case_a}
\end{figure}

For the $X$-checks, consider the six checks on the faces of a single cube, as illustrated in Fig.~\ref{case_a}(a). When these six $X$-checks are multiplied together, the operators on the shared edges cancel out, yielding the identity. This shows that the six $X$-checks on a single cube are not independent, constituting a redundancy. The number of such local redundancies, one per cube, is $L_1 L_2 L_3$. However, these redundancies are not all independent. Any one of them can be expressed as a product of the others, leading to $L_1 L_2 L_3 - 1$ independent relations of this type.

Additional redundancies arise when multiplying all $X$-checks on a plane that spans the entire section of the lattice, as shown in Fig.~\ref{case_a}(b). This also results in the identity. For each of the three independent lattice directions, only one such planar redundancy is independent. The redundancies associated with other planes can be generated by combining the selected planar redundancy with the local cube redundancies described above. Thus, the number of independent planar redundancies is 3.

The total number of independent redundancies among the $X$-checks is therefore $(L_1 L_2 L_3 - 1) + 3$. The total number of $X$-checks is $3L_1L_2L_3$. The number of independent $X$-checks is obtained by subtracting the number of independent redundancies from the total number of checks: $3L_1L_2L_3 - [(L_1 L_2 L_3 - 1) + 3] = 2L_1L_2L_3 - 2$.

Finally, the code dimension $k$ is the total number of physical qubits minus the number of independent checks (both $Z$- and $X$-type). The number of physical qubits is $3L_1L_2L_3$. The number of independent $Z$-checks is $L_1 L_2 L_3 - 1$, and the number of independent $X$-checks is $2L_1L_2L_3 - 2$. Therefore, the code dimension is
\[
k = 3L_1L_2L_3 - [(L_1 L_2 L_3 - 1) + (2L_1L_2L_3 - 2)] = 3.
\]
This confirms that the code dimension is a constant, independent of the system size.

Regarding the logical operators, we note that for an X operator to commute with all $Z$-checks, it must have support on an even number of qubits (i.e., edges) incident to each cube vertex. This condition is satisfied if the $X$-type operators form closed loops in the lattice. However, any $X$-loop in the bulk that is the boundary of a two-dimensional surface can be expressed as a product of the $X$-checks on that surface. Therefore, only non-contractible $X$-loops that traverse the entire lattice, as illustrated in Fig.~\ref{case_a}(c), correspond to logical operators. Among these, there is exactly one independent logical $X$-operator in each spatial direction.

We now consider the dual lattice, where the roles of faces and edges are interchanged. The $X$-checks are located on the edges of the dual lattice, with each check acting on the four faces sharing that edge. For a $Z$-type operator to commute with all $X$-checks, it must form a closed surface in the dual lattice. However, any closed surface that is the boundary of a three-dimensional volume can be written as a product of $Z$-checks. Thus, a logical $Z$-operator corresponds to a non-contractible surface that isn't a boundary and extends to a transversal plane of the dual lattice, as shown in Fig.~\ref{case_a}(d). Here too, there is one independent logical $Z$-operator per direction.

In summary, the length of the logical $X$-operator (a non-contractible loop) is clearly shorter than that of the logical $Z$-operator (a non-contractible surface). Therefore, the code distance of the 3D toric code is given by $\min(L_1, L_2, L_3)$.

\subsection*{Case B}
In Case B, the total number of checks equals the total number of qubits. Consequently, the code dimension is determined directly by the number of redundancies among these checks.

\begin{figure}[t]
    \centering
    \includegraphics[width=\columnwidth]{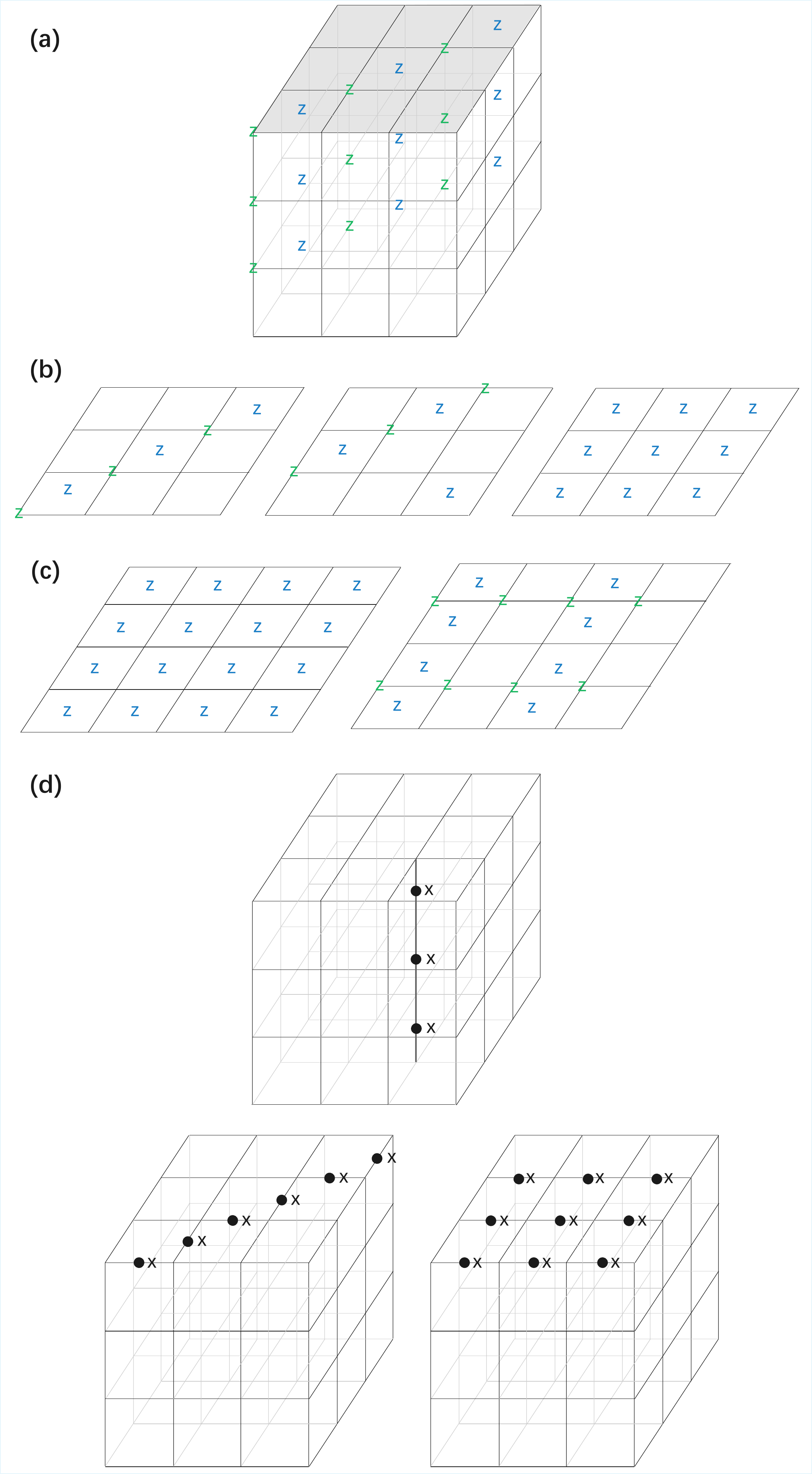}
    \caption{Redundancies of checks and logical operators for Case B. (a) Overall structure of the redundancies among the $Z$-check operators, across all layers on which the checks reside. (b) Examples of the $Z$-check redundancy structure within a single layer for $L_1=L_2=3$, corresponding to the shaded layer in (a). (c) Examples of the $Z$-check redundancy structure within a single layer for $L_1=L_2=4$. (d) Representative logical $X$ operators.}
    \label{case_b}
\end{figure}

A defining characteristic of this configuration is its layered structure along the $z$-direction, where each layer (a plane of constant $z$) contains checks of exclusively one type: either all $Z$-checks or all $X$-checks. To elucidate the origin of the redundancies, we consider a concrete example with $L_1= L_2= L_3 = 3$. Focusing on the qubits on the top layer (as illustrated by the shaded layer in Fig.~\ref{case_b}(a)), we observe that multiplying all $Z$-checks along a diagonal direction forming a closed loop under periodic boundary conditions yields the identity. To achieve a full cancellation that also accounts for the checked qubits in the layers above and below, the product of checks must be taken along the corresponding diagonal loop across \emph{all} $Z$-check layers (layers where $Z$-checks reside on; similarly hereafter) in the lattice. This process exemplifies the general mechanism generating redundancies in the system.

Figure~\ref{case_b}(b) depicts possible redundancy patterns on a single layer (noting that the corresponding checks from all layers must be multiplied together to form a complete redundancy). The first pattern corresponds to the northeast-oriented diagonal loop already described. Shifting this loop of $Z$-checks to the left by one unit, as shown in the second pattern, yields an independent redundancy. Similarly, shifting it to the right by one unit provides a third, independent redundancy. These three possibilities exhaust the independent loops oriented along the northeast direction. An analogous set of three independent loops exists along the northwest direction. However, not all six are independent. Since the product of all northeast loops involves every $Z$-check in the system, and similarly for the northwest loops, a global dependency arises. This implies that only five of these six loop-like redundancies are linearly independent. Furthermore, an additional, distinct redundancy is contributed by a pattern that encompasses all $Z$-checks located at the face centers across the entire plane, as shown in the third figure. Therefore, in total, the $3\times 3\times 3$ system exhibits six independent redundancies.

The presence of a uniform structure along the $z$-direction implies that the code dimension is independent of $L_3$. However, its dependence on $L_1$ and $L_2$ is non-monotonic with respect to the system size. 

A loop traversing in the northeast direction must pass through $2\operatorname{lcm}(L_1, L_2)$ $Z$-checks before closing, since the loop length must be a common multiple of both $L_1$ and $L_2$ to wrap completely around the $x$ and $y$ directions, multiplied by two because each unit cell contains two $Z$-checks of different colors. Given that the total number of $Z$-checks in one layer is $2L_1L_2$, the number of such independent northeast loops is given by
\[
\frac{2L_1L_2}{2\operatorname{lcm}(L_1, L_2)} = \gcd(L_1, L_2).
\]
By also considering the independent northwest loops and the full-plane pattern, the total number of redundant $Z$-checks sums to $2\gcd(L_1, L_2)$.

An important subtlety arises when both $L_1$ and $L_2$ are even, as in the $4\times 4$ example illustrated in Fig.~\ref{case_b}(c). In this case, the redundancy involving all face centers can be generated by combining northwest and northeast loops, and thus it is no longer independent. However, a new independent redundancy pattern emerges under this condition, as shown in the second subfigure, which exists only when both $L_1$ and $L_2$ are even.

Since the $X$-checks and qubits on the half-integer-$z$ layer can be defined by applying a spatial shift to the $Z$-check and qubit positions on the integer-$z$ layer, while simultaneously changing the check type from $Z$ to $X$, the number of redundancies for the $X$-checks is likewise $2\gcd(L_1, L_2)$. Consequently, the total code dimension is given by
\[
4\gcd(L_1, L_2).
\]

For the same reason, the logical operators can be constructed by first deriving the $X$-type logical operators and then obtaining the corresponding $Z$-type logical operators through a spatial shift. 

The first type of $X$ logical operator involves qubits situated at the centers of edges oriented along the $z$-direction, forming a closed loop as depicted in the upper panel of Fig.~\ref{case_b}(d). Each qubit in this operator is acted upon by only two $Z$-stabilizers: those located at the endpoints of its corresponding edge. Since each of these $Z$-stabilizers overlaps with two qubits in the loop, the operator commutes with all stabilizers. The length of such an $X$ logical operator is $L_3$.

The second type of $X$ logical operator is localized within a single $Z$-check layer. To commute with the $Z$-checks in that layer, this operator must form a pattern analogous to the redundancy structures discussed earlier. As illustrated in the second row of Fig.~\ref{case_b}(d), three distinct configurations are possible: northeast-oriented loops (left subfigure) of length $2\operatorname{lcm}(L_1,L_2)$, northwest-oriented loops (not shown) also of length $2\operatorname{lcm}(L_1,L_2)$, and a pattern spanning the entire plane involving only the $X$ operators on the horizontal bonds (right subfigure) with a length of $L_1 L_2$.

In summary, the distance of the code is the smallest of the logical operators:
\[
\min(L_3,{\rm lcm}(L_1,L_2),L_1L_2).
\]
 
\begin{figure}[t]
    \centering
    \includegraphics[width=\columnwidth]{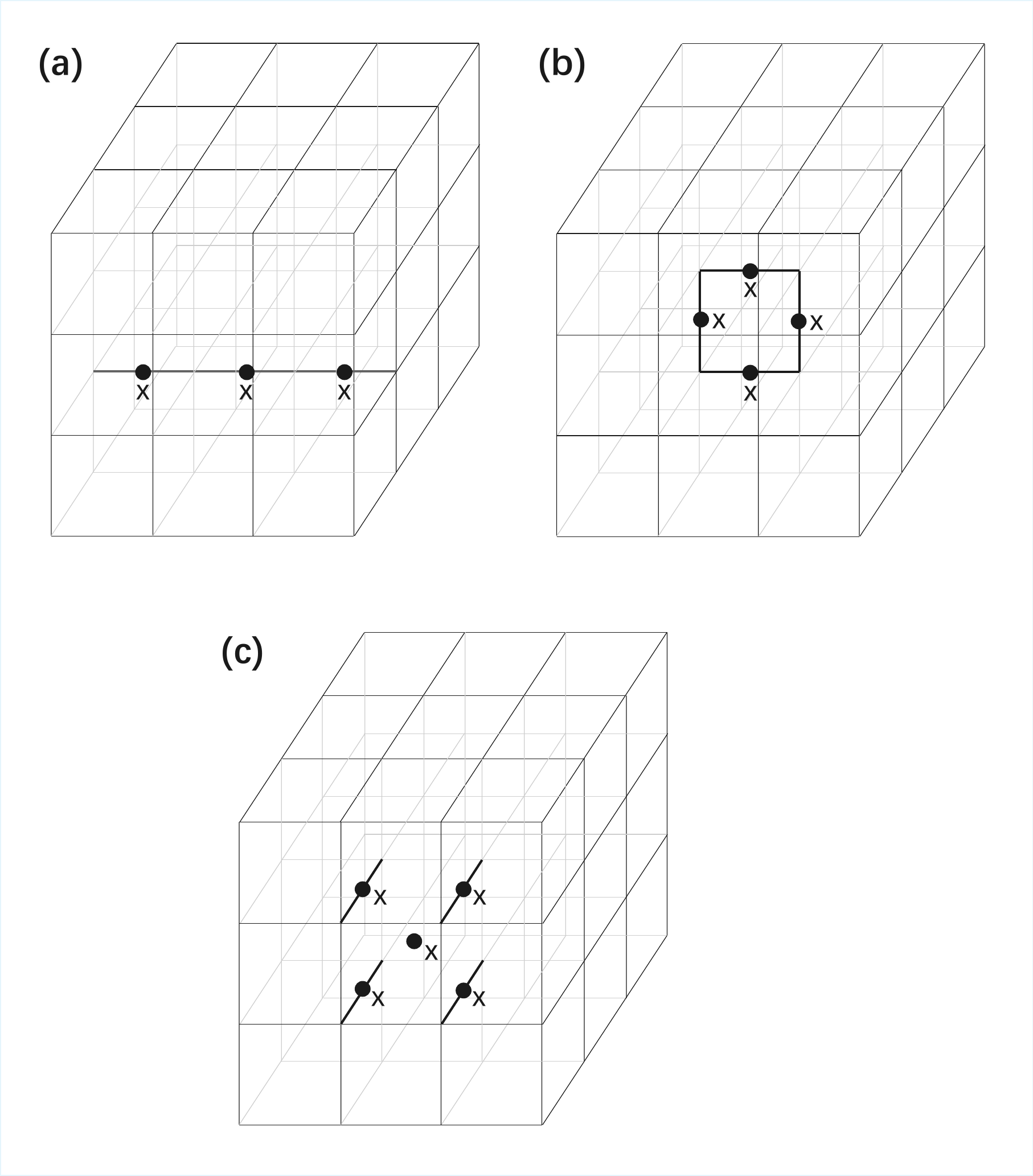}
    \caption{Representative logical $X$ operators for Case C.}
    \label{case_c}
\end{figure}

\subsection*{Case C}

In Case C, the spatial distribution of the $Z$-checks and $X$-checks is identical to that in Case A, with the additional inclusion of qubits at the cube centers and more qubits participating in each check. Under this configuration, multiplying all the $Z$-check operators yields the identity. However, the redundancies among the $X$-checks become too complex to analyze through analytical means. Therefore, the code dimension $k$ is computed numerically. Numerical results indicate that $k$ is positively correlated with the greatest common divisor (GCD) of the lengths of the three classical repetition codes, though the relationship is not as straightforward as in Case B. A rigorous analytical explanation for this observation remains an open question.

Regarding the logical operators, the $X$ operators are believed to have shorter support than the $Z$ operators due to their relatively simple structure. In this setting, the $X$-type operators shown in Fig.~\ref{case_c}(a) remain valid. However, additional local $X$ operators also appear. Our numerical analysis shows that when any pair among $L_1$, $L_2$, and $L_3$ shares a common divisor greater than 1, the length-four loops illustrated in Fig.~\ref{case_c}(b) become logical $X$ operators, since they are no longer stabilizers. In contrast, if no pair shares a common divisor greater than 1, the length-four loops can still be generated by stabilizers, but the configuration in Fig.~\ref{case_c}(c)—comprising four $X$ operators on qubits on the four edges of one cube along the same direction  plus one on the body center—forms a valid $X$ logical operator.

As a result, the code distance is upper-bounded by five. More precisely, it is $\min(L_1, L_2, L_3, 4)$ when a common divisor larger than 1 exists among any pair of $L_1$, $L_2$, and $L_3$, and $\min(L_1, L_2, L_3, 5)$ otherwise. Since the lengths of the logical operators do not scale with the system size, this code is not a practical candidate for quantum error correction.

\subsection*{Case D}
In Case D, the spatial arrangement of checks remains identical to that of Case B. The modification lies in the definition of each check operator, which is extended to act on eight additional qubits from neighboring layers. Specifically, for an $X$-check—which in Case B acts only on the single qubit directly above it in the adjacent $Z$-check plane—we now include the four next-nearest neighbors of that qubit within the same plane. While the redundancies in Case B remain valid in this case, the extension of checks enable more possible structures of redundancies. In Case B, to cancel out on the intermediate $X$-check layers, the same pattern of $Z$-check assemblies must be repeated across all $Z$-check layers. In Case D, however, the extended support allows $Z$-checks within a single layer to cancel out contributions on the upper and lower adjacent planes on themselves, giving rise to new types of redundancies that were previously absent.

\begin{figure}[t]
    \centering
    \includegraphics[width=\columnwidth]{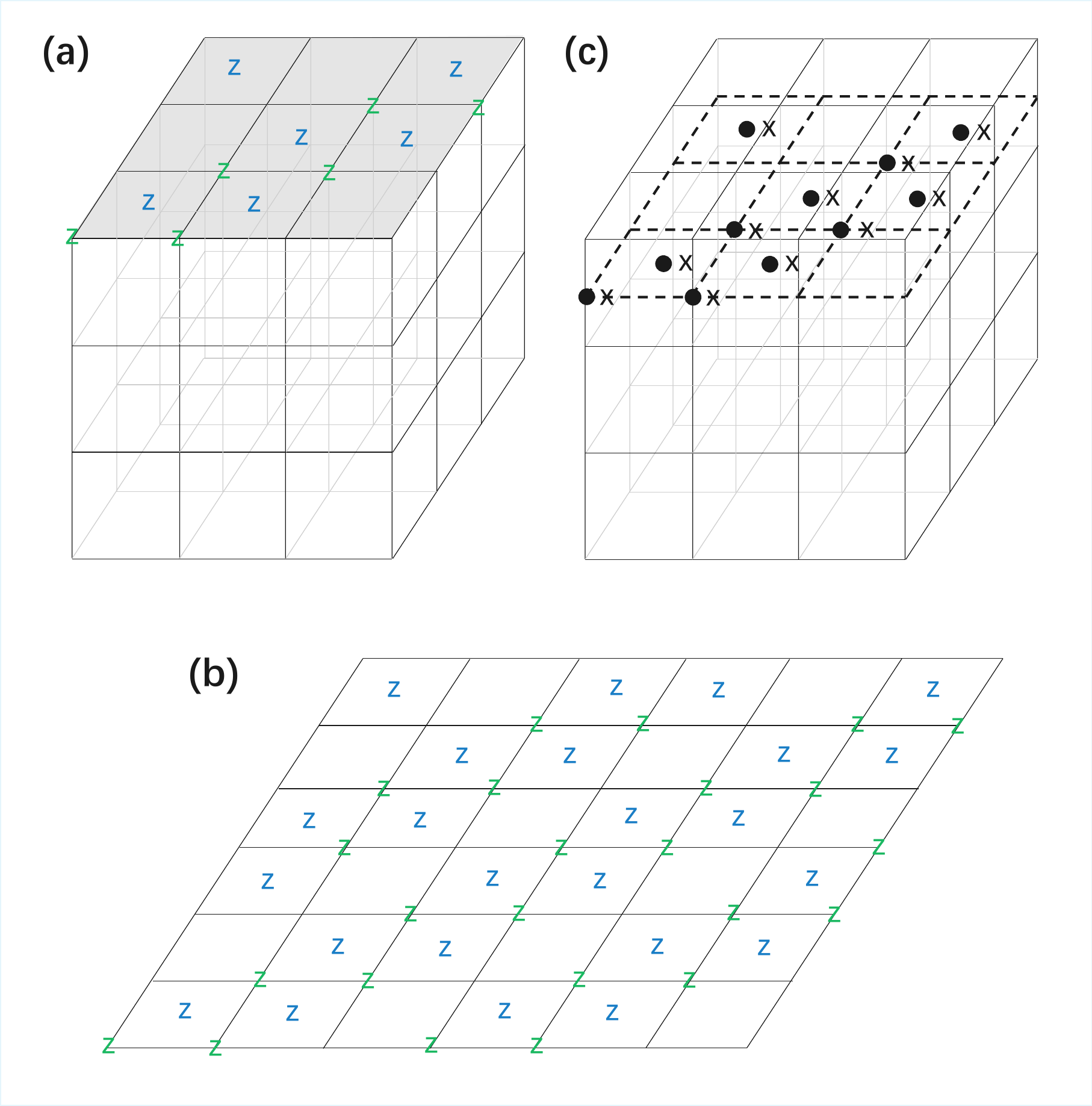}
    \caption{Redundancies of checks and logical operators in Case D. (a) New $Z$-check redundancies for $L_1=L_2=3$, distinct from those in Case B. The shaded plane indicates the layer on which the illustrated checks reside. (b) New $Z$-check redundancies for $L_1=L_2=6$ on one $Z$-Check layer; the pattern shown can occur on any $Z$-check plane in the system. (c) An additional representative logical $X$ operator. The dotted lines indicate the plane on which the operator is located.}
    \label{case_d}
\end{figure}

Novel redundancy patterns emerge when both $L_1$ and $L_2$ are multiples of three. As illustrated in Fig.~\ref{case_d}(a), these can involve $Z$-checks reside only on a single layer. To achieve cancellation within this layer, the new redundancy must correspond to a combination of previously established redundancies in Case B restricted to the plane. The correct configuration involves two adjacent northeast-oriented loops, which mutually cancel on the upper and lower layers through boundary conditions perpendicular to the loops. Generalization to larger planes is straightforward: including two out of every three neighboring northwest loops suffices to form a valid redundancy, see Fig.~\ref{case_d}(b). This configuration is feasible if and only if both $L_1$ and $L_2$ are divisible by three.

To determine the number of independent single-layer redundancies, we first note that shifting the pattern by one unit along right direction produces a distinct configuration. However, further shifts yield redundant combinations expressible as linear combinations of the previous ones. Considering both northeast and northwest orientations, each layer supports four independent redundancy patterns. Although the system contains $L_3$ $Z$-check layers, only $4 \times (L_3 - 1)$ of these redundancies are independent, as any single-layer redundancy can be constructed from all-layer redundancies combined with single-layer redundancies from other layers.

By symmetry, similar considerations apply to $X$-check redundancies. Consequently, the code dimension acquires an additional term that scales nearly linearly with the third dimension:

\[
k = 4\gcd(L_1,L_2) + \alpha(L_3-1),
\]
where $\alpha = 8$ if 3 divides $L_1$ and $L_2 $; otherwise $\alpha = 0$.

Along with the increase in code dimension, new logical operators can be defined when $L_1$ and $L_2$ are divisible by three. A new type of $X$ logical operator appears on the $X$-check layers, which is valid because it now overlaps an even number of times with the support of the $Z$-checks on neighboring layers. The pattern of these operators is similar to that of the single-layer redundancies, requiring the inclusion of two out of every three neighboring northeast- or northwest-oriented loops, with an example shown in Fig.~\ref{case_d}(c). Still, the original logical operators from Case B remain valid in this case. Since the new logical operators are necessarily longer than the shortest one from Case B, the code distance remains the same as in Case B, given by:

\[
\min\left(L_3, \ \mathrm{lcm}(L_1, L_2), \ L_1 L_2\right).
\]